\documentclass[
aps,
12pt,%
final,%
notitlepage,%
oneside,%
onecolumn,%
superscriptaddress,%
noshowpacs,%
centertags]%
{revtex4}
\usepackage{graphicx}

\begin{document}
\selectlanguage{english} 
\title{Vanishing of electron pair recession at central impact}
\author{\firstname{Nataliia O.}~\surname{Cherkashyna}}
\email{ncherkashyna@ukr.net}
\affiliation{National Shevchenko University of Kyiv, Theoretical Physics Department}%

\author{\firstname{Constantin V.} ~\surname{Usenko}}
\email{usenko@univ.kiev.ua}
\affiliation{National Shevchenko University of Kyiv, Theoretical Physics Department}%

\begin{abstract}

Identity of electrons leads to description of their states 
by symmetrical or anti-symmetrical combination of free coherent states.

Due to the coordinate uncertainty potential energy of the Coulomb repulsing 
is limited from above and so when energy of electrons is large enough, electrons go through each other, without noticing one another. 

We show existence of set of coherent states for which wave packages recession vanish - electrons remain close regardless of Coulomb repulsion.
\end{abstract}

\maketitle
\section{Introduction}
Taking into consideration the identity of electrons we describe their state 
by symmetrical or anti-symmetrical combination of free coherent states 
\cite{npyt,sag,pfff,vacjpued}. These states have a lot of essentially quantum pecularities such as entanglement \cite{gafi,lgz,llv,bekvt}.

Due to the coordinate uncertainty potential energy of the coulomb repulsing 
is limited from above. That is why two types of classical interpretation of 
electron interaction are possible:
\begin{itemize}
\item
Classical scattering of an electron on another one. Electrons approach a 
certain distance, but due to the Coulomb repulsion they scatter in opposite 
directions.
\item
If energy of electrons is large enough, electrons go through each other, 
without noticing one another.
\end{itemize}
Two types of classic description of results of an impact take place 
for electrons with parallel and with anti-parallel spins as well.

In these two cases, due to recession of centers of packages of waves the 
entanglement of the state (both spatial and spin) remains constant. For the 
case of antiparallel mutual orientation of electron spins there is a certain 
range of relative momentum values at which recession of the centers of wave 
packages vanishes. Instead, spreading of wave packages takes place. The 
observed characteristic of the state is the quadrupole moment of the 
electric field produced by the pair of charges. In this case the spatial 
entanglement of the states vanishes simultaneously with packages spreading, 
in contrast to spin entanglement. Vanishing of recession makes evidence of 
the fact, that electrons are in the almost identical states long enough.

Will examine the pair of coherent electrons in the center-of-mass system, 
$r_1 = \frac{R}{2} + r$, $r_2 = \frac{R}{2} - r$, $p_1 = \frac{P}{2} + p$, 
$p_2 = \frac{P}{2} - p$. The Hamiltonian of this system is: 

\begin{equation}
	H = \frac{p_1^2 }{2m} + \frac{p_2^2 }{2m} + \frac{e_0^2 }{\left| \vec 
	{r}_1 - \vec {r}_2\right|} 
	= \frac{P^2}{4m} + \frac{p^2}{m}+ \frac{e_0^2 }{\left|\vec{r}_1 - \vec{r}_2 \right|}.
\label{ham1}
\end{equation}

Hamiltonian of coherent electron pair is divided on Hamiltonian of free 
motion of center of mass of electron pair 
$H_c = \frac{\vec {P}^2}{4m}
$
and Hamiltonian of relative motion
\begin{equation}
\label{ham2}
H_r = \frac{\vec {p}^2}{m} + \frac{e_0^2 }{\left| \vec{r} \right|},
\end{equation}

which is similar to the Hamiltonian of electron in the atom of hydrogen.

\section{Wave function of coherent electron pair}
Studying the electromagnetic field of coherent electron, (and also 
electromagnetic field of coherent electron pair), we deal with the problem on effect
 of coordinate and momentum uncertainties of free particle. Plane wave is not 
good instrument for description of the coherent electron state, because a 
flat wave is characterized a zero momentum uncertainty, and coordinate 
uncertainty tends to infinity. The same concerns to the functions, localized 
in space, that look like: $\delta (x - x_0 (t))$. That is why we consider 
the wave function of coherent electron as Gaussian superposition of plane 
waves. Such kinds of probability distribution are characterized by the 
coordinate and momentum uncertainties $\sigma _x $ and$\sigma _p $. For the 
free particle, momentum uncertainty is constant,$\sigma _p = const$, but 
coordinate uncertainty depends on time. 
\subsection{ Wave function of single coherent electron}
Taking into consideration the aforesaid, we can write the wave 
function of coherent electron:

\begin{equation}
\label{eq3}
\Psi (\vec {r},\vec {r}_0 ,\vec {p}_0 ) = \frac{1}{\left( {\sqrt {\sigma 
\sqrt {2\pi (1 + \omega ^2(t - t_0 )^2)} } } \right)^3}\exp \left( { - 
\frac{\left( {\vec {r} - \vec {r}_0 - \frac{\vec {p}_0 }{m}(t - t_0 )} 
\right)^2}{4\sigma ^2(1 + \omega ^2(t - t_0 )^2)} + i\frac{\vec {p}_0 \vec 
{r}}{\hbar }} \right),
\end{equation}
where $\sigma $ is coordinate uncertainty, $r_0 $- an average value of the 
coordinate, $p_0 $- an average value of the momentum, $t_0 $- is a 
culmination moment (moment of time, when correlation between the coordinate 
and momentum is absent), $m$- electron mass. Taking into consideration the 
uncertainty relation, which takes a minimum value in the moment of 
culmination, we can write down: 
\begin{equation}
\label{eq4}
\sigma _x = \frac{\hbar }{2\sigma _p }\sqrt {1 + \omega ^2\left( {t - t_0 } 
\right)^2} 
\end{equation}
whence $\sigma _r \sigma _p = \frac{\hbar }{2}\sqrt {1 + \omega ^2\left( {t 
- t_0 } \right)^2} $.
In the moment of culmination this expression takes a minimum value  
$
\sigma _r \sigma _p =\hbar/2.
$

As it is obvious from (\ref{eq4}), the dependence of the wave package on time 
provides package spreading with time.

The kinetic energy of coherent electron:
\begin{equation}
\label{eq5}
T = \frac{p_0^2 }{2m} + \frac{3\hbar }{4\sigma ^2m},
\end{equation}
where the first item describes the energy of quasi-classical motion of wave 
package center, the second item is conditioned by momentum uncertainty. 

\subsection{1. 2. Wave function of coherent electron pair}
Wave function of coherent electron pair is an anti-symmetrical (in the case 
of parallel mutual spin orientation) or symmetrical (in the case of 
anti-parallel mutual spin orientation) combinations of wave functions of 
every electron (according to Pauli principle):
\begin{equation}
\label{eq6}
\Psi (\vec {r}_1 ,\vec {r}_2 ) = \frac{\Psi _1 (\vec {r}_2 )\Psi _2 (\vec 
{r}_1 )\pm \Psi _1 (\vec {r}_1 )\Psi _2 (\vec {r}_2 )}{\sqrt {1\pm N^2} }.
\end{equation}
$N$ is an overlapping integral, which is defined is such a way:
\begin{equation}
\label{eq7}
N = \int {\Psi _1^\ast (r)} \Psi _2 (r)d^3r = \left\langle {\Psi _1 (r)} 
\mathrel{\left| {\vphantom {{\Psi _1 (r)} {\Psi _2 (r)}}} \right. 
\kern-\nulldelimiterspace} {\Psi _2 (r)} \right\rangle .
\end{equation}
Overlapping integral is not equal to zero because wave functions are not 
orthogonal.

In the center-of-mass system average values of coordinate sum and momentum sum are 
equal to zero. 

The parameters $\vec {r}_0 $ and $\vec {p}_0 $ have to take values of 
relative coordinate and relative momentum correspondingly. But these 
parameters should not be considered as average values of coordinate and 
momentum, because these values are equal to zero due to identity of 
electrons. 

We consider coherent electron pair at $t = t_0 $. The 
wave function is:
\begin{equation}
\label{eq8}
\begin{array}{l}
 \Psi (\vec {r}_1 ,\vec {r}_2 ) = \frac{1}{\sqrt {1\pm N^2} }\frac{1}{\left( 
{\sigma \sqrt {2\pi } } \right)^3} \cdot \\ 
 \cdot \left[ {\exp \left( { - \frac{\left( {\vec {r}_1 - \vec {r}_0 } 
\right)^2}{4\sigma ^2} - \frac{\left( {\vec {r}_2 + \vec {r}_0 } 
\right)^2}{4\sigma ^2} + i\frac{\vec {p}_0 (\vec {r}_1 - \vec {r}_2 )}{\hbar 
}} \right)} \right.\\ \pm \left. {\exp \left( { - \frac{\left( {\vec {r}_2 - 
\vec {r}_0 } \right)^2}{4\sigma ^2} - \frac{\left( {\vec {r}_1 + \vec {r}_0 
} \right)^2}{4\sigma ^2} + i\frac{\vec {p}_0 (\vec {r}_1 - \vec {r}_2 
)}{\hbar }} \right)} \right] \\ 
 \end{array}
\end{equation}
Taking into consideration the wave function, the average of \ref{ham2} is \cite{hnat}: 
\begin{equation}
\label{eq88}
\begin{array}{l}
\left\langle H \right\rangle = \frac{p_0^2 }{m} + \frac{3\sigma ^2}{m}
 + \frac{\sigma ^2\left| {\frac{2\sigma }{\hbar }r_0 + \frac{i}{\sigma }p_0 } 
\right|^2\exp \left( { - \left| {\frac{2\sigma }{\hbar }r_0 + 
\frac{i}{\sigma }p_0 } \right|^2} \right)}{m\left( {1 + \exp \left( { - 
\left| {\frac{2\sigma }{\hbar }r_0 + \frac{i}{\sigma }p_0 } \right|^2} 
\right)} \right)} \\
+ e^2\frac{\frac{1}{\left| R \right|}erf\left( 
{\frac{\sigma \left| R \right|}{\hbar \sqrt {1 + \omega ^2t^2} }} \right) - 
\frac{i}{\left| L \right|}erf\left( {\frac{\sigma \left| L \right|}{\hbar 
\sqrt {1 + \omega ^2t^2} }} \right)\left| J \right|^2}{1 + \left| J 
\right|^2}
\end{array}
\end{equation}
where $\left| R \right| = \frac{1}{2}\frac{1}{r(t) + 
\frac{t}{m}p(t)}$, $\left| L \right| = \frac{\hbar }{\sigma ^2}\vec {p}(t) - 
2\omega t\vec {r}(t)$, $\left| J \right|^2 = \exp (\left( { - \left| 
{\frac{2\sigma }{\hbar }r(t) + \frac{i}{2}p(t)} \right|} \right)$
The average value of Hamiltonian of coherent electron pair depends on time. 
This dependence is determined by the fact that correlation between relative 
coordinate and relative momentum is absent in the initial moment of time 
only (moment of culmination).

\section{Equations of motion}

Using the method of integration by trajectories, Klauder has proved that one 
can obtain classical Hamiltonian equations of motion considering the average 
value of quantum system Hamiltonian as the Hamiltonian of classical system 
\cite{kla}. 

These equations go to classic equations of free motion of center of mass 
and equations of relative motion
\begin{equation}
\label{eq9}
\frac{d\vec {p}}{dt} = - \frac{\partial \left\langle H \right\rangle 
}{\partial \vec {r}},
\frac{d\vec {r}}{dt} = \frac{\partial \left\langle H \right\rangle 
}{\partial \vec {p}}
\end{equation}
Thus, in the Hamilton equations one should use relative coordinate and 
momentum. Equations of motion:
\begin{equation}
\label{eq10}
\frac{dp}{dt} = - \frac{\partial E}{\partial r} \quad ,
\quad
\frac{dr}{dt} = \frac{\partial E}{\partial p},
\end{equation}
where $E = \left\langle H \right\rangle $.

The average value of Hamiltonian of coherent electron pair depends on time. 
This dependence is determined by the fact that correlation between relative 
coordinate and relative momentum is absent in the initial moment of time 
only (moment of culmination).

The average value of energy of the system depends on time. This dependence 
is determined by the special quantum properties, first of all, by the 
effect of relative coordinate and momentum uncertainties. Even if the 
average values of coordinate and momentum are equal to zero, the average 
value of Hamiltonian depends on time due to spreading of wave 
package. 

\section{Central impact of coherent electrons}

\subsection{Quasi-classical model of the central impact }

Due to the coordinate uncertainty potential energy of the coulomb repulsing 
is limited from above. That is why two types of classical interpretation of 
electron interaction are possible. 

\begin{enumerate}
	\item 
Scattering (same to the classical one) of an electron on another one. 
Electrons approach a certain distance, but due to the Coulomb repulsion they 
scatter in opposite directions (classical-like). 
	\item 
If energy of electrons is large enough, electrons go through each other, 
without noticing one another (tunneling-like).
\end{enumerate}

Taking into consideration the principle of identity we find these two cases 
to be physically indistinguishable. We can not distinguish, which electron 
goes in certain direction. We can not find, whether they continue motion in 
the same directions, as before the impact (it means electrons fly over 
through each other), or in the opposite directions (it means, Coulomb 
repulsion took place).

We can calculate time $t$ electrons need to return to an initial distance. 
The traveltime differs according to the type of impact.

In classical approach to consideration of forward scattering, the falling 
electron stops, and the electron, which is the target, after scattering goes 
with the same speed the falling electron has had before scattering. Taking 
into consideration the principle of identity, we can not distinguish between the 
falling electron and the target electron. At the same time it is supposed, that 
Coulomb interaction between electrons remains and the potential interaction 
energy tends to infinity. 

\begin{figure}[h]
\setcaptionmargin{5mm}
\onelinecaptionsfalse %
\includegraphics[width=2in]{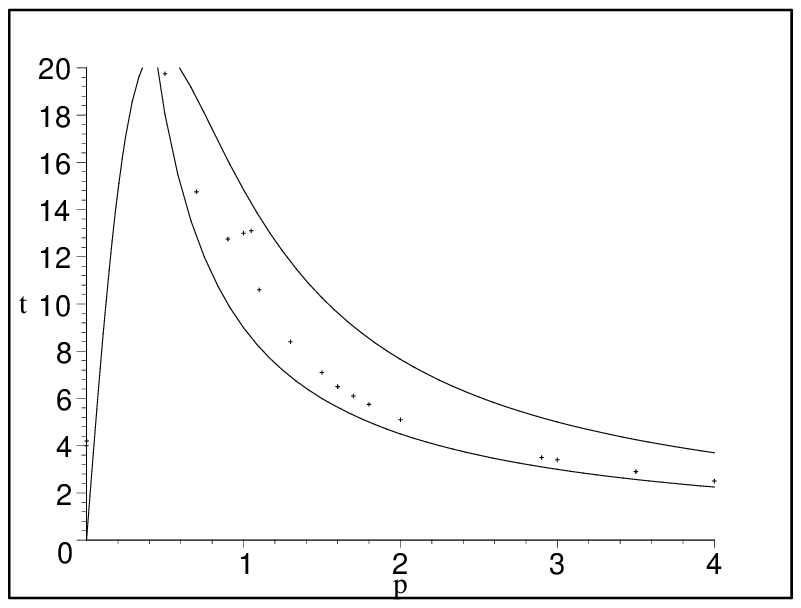}%
\includegraphics[width=2in]{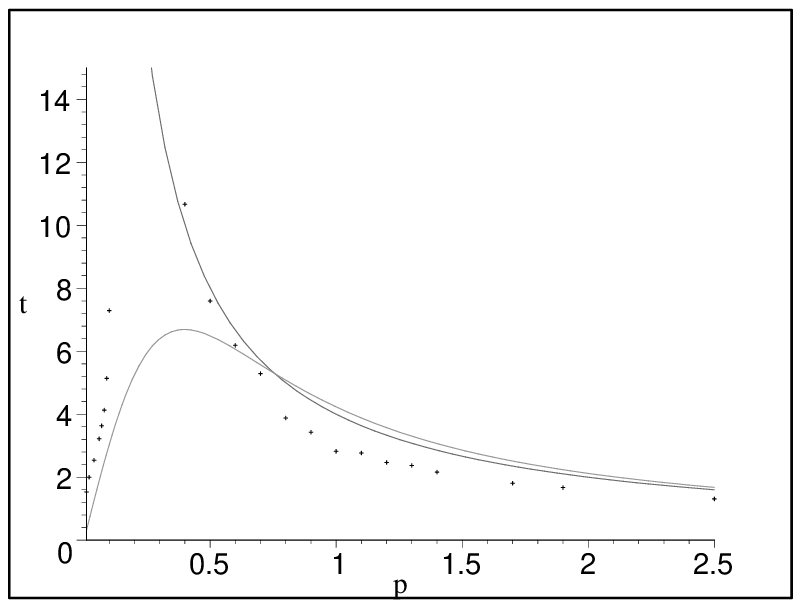}
\captionstyle{normal}\caption{The traveltime dependence on relative momentum values.\\ Left -- case of parallel mutual spin orientation, right -- anti-parallel one.\\ Curve 1 corresponds to electrons travelflying over through each other; time is inversely proportional to the momentum; Curve 2 corresponds to classical Coulomb 
collision. Curve 3 corresponds to two coherent electrons interaction.
}
\label{time_dep}
\end{figure}

Dependence of traveltime on initial momentum is shown in figure \ref{time_dep}.

Studying the central impact of coherent electrons, we can outline two types 
of traveltime behavior. For the first traveltime is definite at the any 
value of relative momentum. It takes place for the electrons with parallel 
mutual spin orientation, and the special case, for the electrons with 
anti-parallel mutual spin orientation as well, except the electrons, which 
have the relative momentum in the certain range. For the special case the 
traveltime is indefinitely large. It means, electrons do not return to the 
initial distance long enough.

\subsection{Quadrupole moment of electric field}

Instead of attempt to build an operator, for which parameters$r_0 $, $p_0 $ (or 
their combination) are the eigenvalues, we study the quadruple moment 
dependence on these values. The quadrupole moment of electric field, 
produced by the pair of charge, is a well determined physical quantity. 

Quadrupole moment for the system of classical charges is described by the 
following:

\begin{equation}
\label{eq11}
D_{\alpha \beta } = \sum\limits_i {q_i } \left( {3x_i^\alpha x_i^\beta - 
r_i^2 \delta _{\alpha \beta } } \right)
\end{equation}

For the case of optional mutual orientation of vectors $\vec {p}_0 $ and 
$\vec {r}$ ($\vec {r} = (x,y,z)$, $\vec {p}_0 = (p_{0x} ,0,p_{0z} )$, $\vec 
{r}_0 = (0,0,r_0 )$), we calculate the components of tensor of quadrupole 
moment:

\begin{equation}
\label{eq12}
D = \left( {{\begin{array}{*{20}c}
 {D_{xx} } \hfill & 0 \hfill & {D_{xz} } \hfill \\
 0 \hfill & {D_{yy} } \hfill & 0 \hfill \\
 {D_{zx} } \hfill & 0 \hfill & {D_{zz} } \hfill \\
\end{array} }} \right),
\end{equation}

here$D_{xz} = D_{zx} $.

Nonzero components of the tensor are defined:

\begin{equation}
\label{eq13}
\begin{array}{l}
D_{xx} = \int {\rho \left( {x,y,z} \right)\left( {2x^2 - y^2 - z^2} \right)} 
dxdydz,\\
D_{yy} = \int {\rho \left( {x,y,z} \right)\left( {2y^2 - x^2 - z^2} \right)} 
dxdydz,\\
D_{zz} = \int {\rho \left( {x,y,z} \right)\left( {2z^2 - x^2 - y^2} \right)} 
dxdydz,\\
D_{xz} = D_{zx} = \int {\rho \left( {x,y,z} \right)xz} dxdydz,
\end{array}
\end{equation}
where $\rho (x,y,z)$ is density of probability. 

Components of tensor are: 
\begin{equation}
\label{eq17}
\begin{array}{l}
D_{xx} = \frac{2N^2\left( {\frac{4\sigma ^4}{\hbar ^2}\left( {p_{0z}^2 - 
2p_{0x}^2 } \right)} \right) - 2r_0^2 }{1\pm N^2},\\
D_{yy} = \frac{2N^2\left( {\frac{4\sigma ^4}{\hbar ^2}\left( {p_{0z}^2 + 
p_{0x}^2 } \right)} \right) - 2r_0^2 }{1\pm N^2},\\
D_{zz} = \frac{2N^2\left( {\frac{4\sigma ^4}{\hbar ^2}\left( {p_{0x}^2 - 
2p_{0z}^2 } \right)} \right) + 4r_0^2 }{1\pm N^2},\\
D_{xz} = D_{zx} = \frac{ - 12N^2\frac{\sigma ^2p_{0x} p_{0z} }{\hbar 
^2}}{1\pm N^2}.
\end{array}
\end{equation}

Let us consider the case$N \to 0$. It is possible if $4\left(p_{0x}^2 + 
p_{0z}^2\right)\sigma^2/\hbar^2 \gg 1$ , and $ r_0^2 /\sigma ^2 \gg 
1$, it means $r_0 \gg \sigma $ and $(p_{0x}^2 + p_{0z}^2 )\gg \left(\hbar /2\sigma \right)^2 $. 

Under those restrictions one can determine $r_0 $:

\begin{equation}
\label{eq21}
r_0 = \sqrt { - \frac{D_{xx} }{2}} = \sqrt { - \frac{D_{yy} }{2}} = 
\frac{1}{2}\sqrt {D_{zz} } .
\end{equation}

Let us consider the case of$N \to 1$, to determine $p_{0x} $ and $p_{0z} 
$values. It is possible in the case of$\frac{4(p_{0x}^2 + p_{0z}^2 )\sigma ^2}{\hbar 
^2} < < 1$ and $ - \frac{r_0^2 }{\sigma ^2} < < 1$, which means $(p_{0x}^2 + 
p_{0z}^2 ) < < \frac{\hbar }{4\sigma }$ and $r_0 < < \sigma $. One can 
get following expressions:

\begin{equation}
\label{eq22}
p_{0x} = \frac{\hbar }{2\sigma ^2}\sqrt { - \frac{D_{zz} + 2D_{xx} }{3}} ,
\end{equation}

\begin{equation}
\label{eq23}
p_{0z} = \frac{1}{\hbar }\sqrt {\frac{1}{3}} \frac{D_{xz} }{\sqrt { - 
(D_{zz} + 2D_{xx} )} }.
\end{equation}

The quadrupole moment: $D = D_{zz} \cos ^2(\vartheta ) + \left( {D_{xx} \cos 
^2(\varphi ) + D_{yy} \sin ^2(\varphi )} \right)\sin ^2(\vartheta ) + 
2D_{xz} \cos (\vartheta )\sin (\vartheta )\cos (\varphi )$ (26)

\begin{figure}[h]
\setcaptionmargin{5mm}
\onelinecaptionsfalse %
\includegraphics[width=2in]{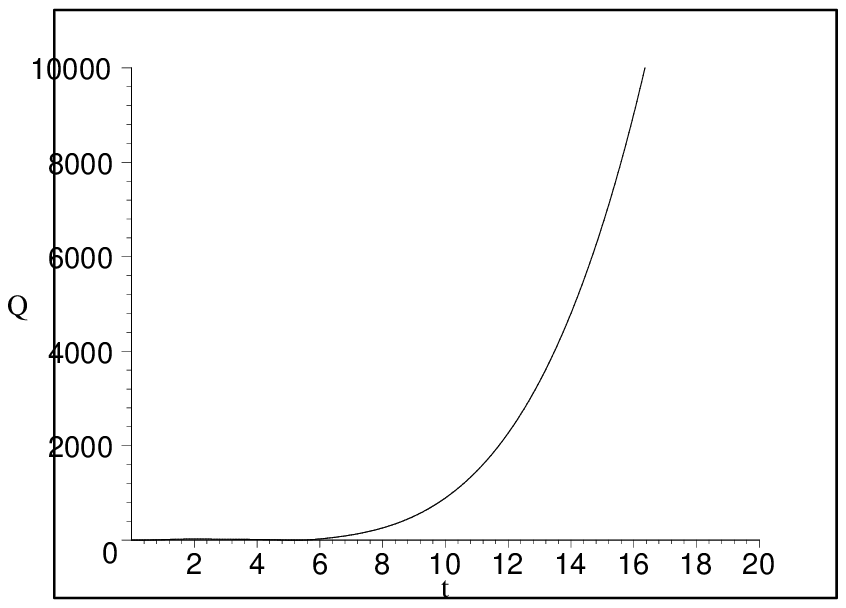}
\includegraphics[width=2in]{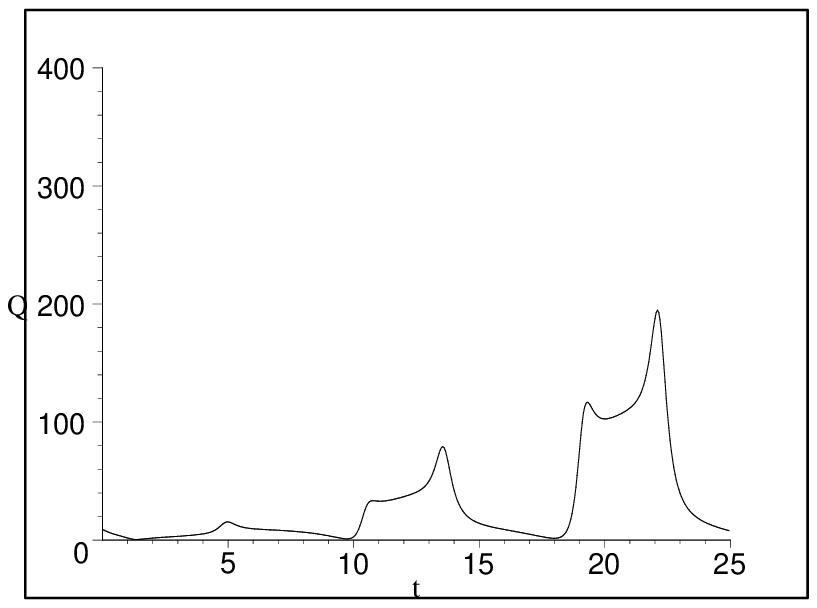}
\captionstyle{normal}\caption{The dependence of quadrupole moment on time: left -- typical case, right -- special one.}
\label{quad_mom}
\end{figure}

The dependence of quadrupole moment on time is shown in figure \ref{quad_mom}. It is monotone for common case and periodic for the special 
one. 

\subsection{Recession and spreading of wave packages}
\begin{figure}[h]
\setcaptionmargin{5mm}
\onelinecaptionsfalse %
\includegraphics[width=2in]{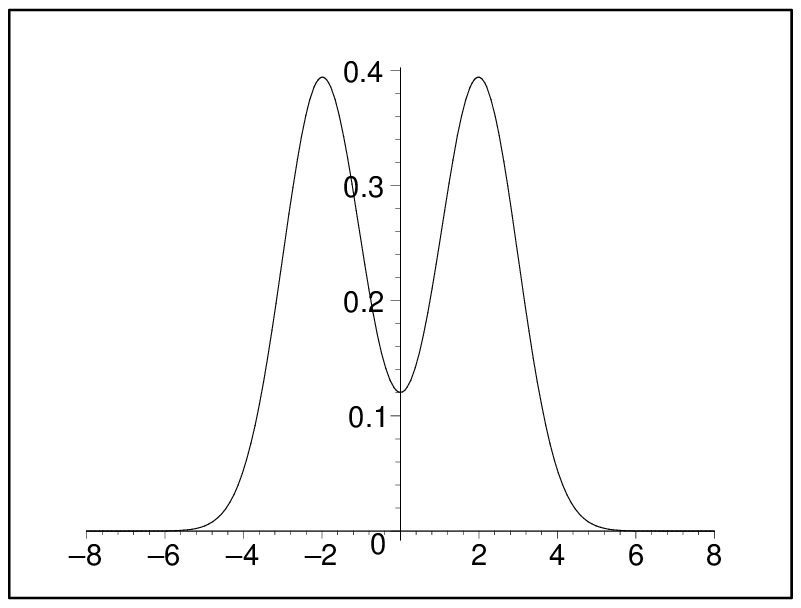}%
\includegraphics[width=2in]{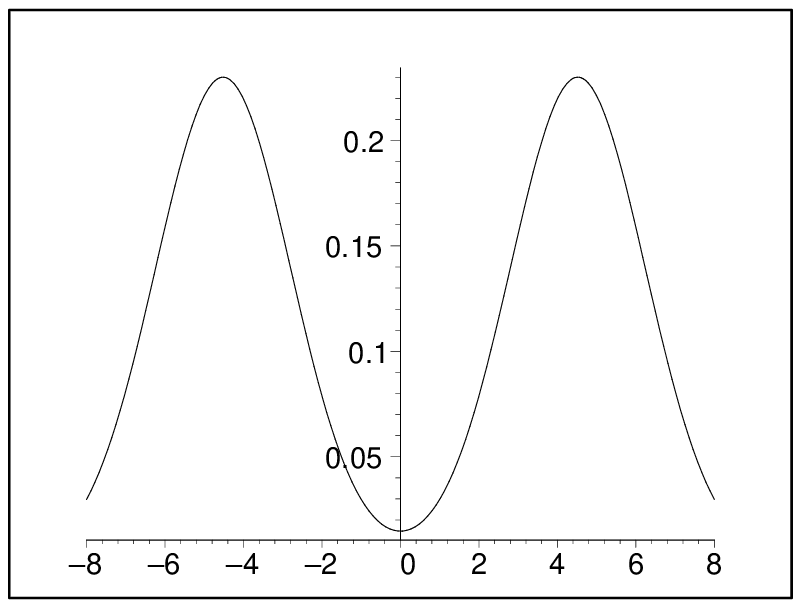}%
\includegraphics[width=2in]{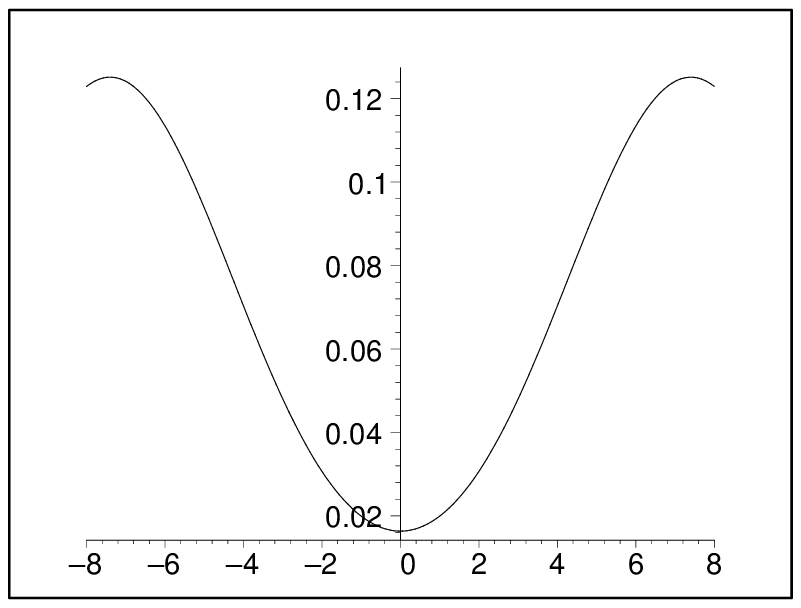}
\captionstyle{normal}\caption{Change of wave function of coherent electron pair is represented for a case of recession of centers of wave packages quicker than their spreading.}
\label{waves1}
\end{figure}

For the typical case of coherent electrons, the recession of centers of wave 
packages is quicker than spreading of packages as it is shown in figure \ref{waves1}.

\begin{figure}[h]
\setcaptionmargin{5mm}
\onelinecaptionsfalse %
\includegraphics[width=2in]{wf_1.eps}%
\includegraphics[width=2in]{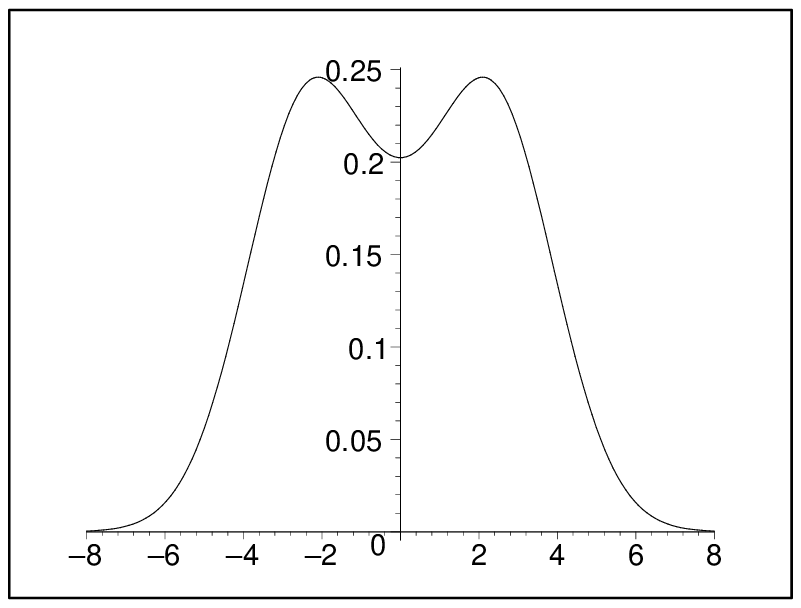}%
\includegraphics[width=2in]{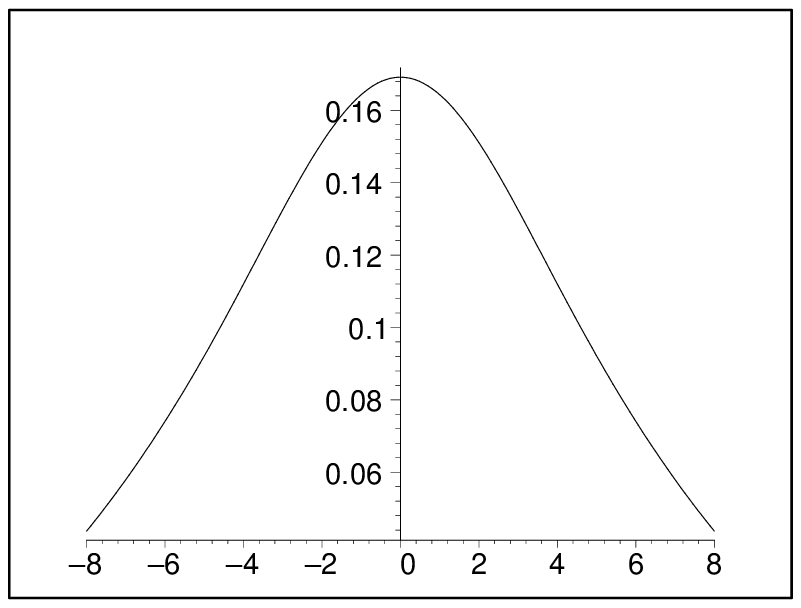}
\captionstyle{normal}\caption{Change of wave function of coherent electron pair is represented, recession of centers of wave packages vanishes, instead spreading of wave packages 
takes place.}
\label{waves2}
\end{figure}

For the special case (figure \ref{waves2}), the recession of centers of wave packages is 
more slow, then their spreading. In fact, the recession of the centers of 
wave packages vanishes with time. Spreading of wave packages takes place. 
The dependence of quadrupole moment of electric field, produced by the 
electron pair, on time is periodic in the special case.

\section{Conclusions}
Coordinate and momentum uncertainty leads to a lot of especially quantum peculiarities of electron pair Coulomb repulsing.  Here we have shown existence of set of coherent states for which wave packages recession vanish - electrons remain close regardless of Coulomb repulsion.

\newpage


\end{document}